\documentclass[a4paper,11pt]{article}
\usepackage{pos}
\usepackage{caption}
\usepackage{subcaption}

\renewcommand{\logo}{\relax}

\title{Higgs pair production at NNLO}

\author*[a]{Joshua Davies}

\affiliation[a]{Department of Mathematical Sciences, University of Liverpool,\\
Liverpool, L69 3BX, UK}

\emailAdd{j.o.davies@liverpool.ac.uk}

\abstract{In this talk I report on progress towards the computation of the next-to-next-to-leading order
virtual corrections to Higgs boson pair production in gluon fusion, in an expansion around the forward-scattering
limit. The results of this computation will provide an estimate of the size and uncertainty of the
three-loop virtual corrections in a comparatively much larger region of phase space than the results available
in the literature, which are based on an expansion in the limit of a large top quark mass.}

\FullConference{Loops and Legs in Quantum Field Theory (LL2024)\\
 14-19, April, 2024\\
Wittenberg, Germany\\}

\begin{document}
\hfill LTH 1376
\maketitle

\section{Introduction}
A primary objective of CERN's Large Hadron Collider (LHC) is to understand the scalar sector of the
Standard Model (SM) by measuring the behaviour of the Higgs boson, which was discovered in 2012
\cite{ATLAS:2012yve,CMS:2012qbp}. In the SM the Higgs boson has cubic and quartic self couplings,
with a strength determined by the coupling constant $\lambda$. Its value is determined by other, known
quantitites: $\lambda = m_H^2/(2v^2)$, where $m_H$ is the mass of the Higgs boson and $v$ is its vacuum
expectation value. An independent experimental measurement of $\lambda$ is required in order to determine if the
structure of the SM's scalar sector is indeed observed in nature.

To make such a measurement, we must study a process which depends on $\lambda$. The most promising such
process is the production of a pair of Higgs bosons; the large gluon luminosity at
the LHC makes gluon fusion the dominant production mechanism by approximately an order of magnitude. The leading
order (LO) cross section for this process has been known analytically for several decades \cite{Glover:1987nx,Plehn:1996wb}
however at next-to-leading order (NLO) and beyond, no fully analytic results are known. Particularly
at NLO, the amplitude is nonetheless well studied via numerical evaluation \cite{Borowka:2016ehy,Borowka:2016ypz,Baglio:2018lrj}
or via expansion in a variety of kinematic limits, including the large top quark mass limit \cite{Dawson:1998py,Grigo:2013rya,Degrassi:2016vss},
the high-energy limit \cite{Davies:2018ood,Davies:2018qvx,Davies:2019dfy}, the small Higgs transverse momentum limit
\cite{Bonciani:2018omm}, the small-$t$ limit \cite{Davies:2023vmj}, around the top quark pair threshold \cite{Grober:2017uho},
and an expansion for small Higgs boson mass followed by numerical evaluation \cite{Xu:2018eos,Wang:2020nnr}.
Since such expansions are valid only in a particular region of phase space, the combination of different
expansions can produce an approximation which is valid everywhere. Such combinations have been investigated in
\cite{Davies:2019dfy,Bellafronte:2022jmo,Bagnaschi:2023rbx,Davies:2023vmj}.
These various works show that the NLO correction is large, approximately the same size as the LO
amplitude. Additionally, it has a large uncertainty due to its dependence on the renormalization
scheme and scale used for the top quark mass \cite{Baglio:2018lrj,Bagnaschi:2023rbx}.

At next-to-next-to leading order (NNLO) comparatively little is known; an approximation has been made by
combining the heavy-top limit with numerically evaluated real radiation contributions \cite{Grazzini:2018bsd},
and the large top quark mass expansion of the virtual \cite{deFlorian:2013jea,Grigo:2015dia,Davies:2019djw}
and real \cite{Davies:2019esq,Davies:2019xzc} contributions are available.
In these proceedings we review the small-$t$ expansion used by Ref.~\cite{Davies:2023vmj} at NLO and
Ref.~\cite{Davies:2023obx} at NNLO, and summarize the status of ongoing work.

\section{Notation}
\begin{figure}[b]
	\centering
	\begin{subfigure}{0.3\textwidth}
		\includegraphics[width=\linewidth]{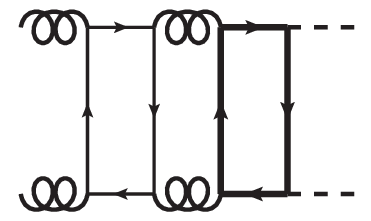}
		\captionsetup{justification=centering}
		\caption{\textit{One light and one heavy quark\\loop: ``$n_l$ diagrams''.}}
		\label{fig:dias:nl}
	\end{subfigure}
	\begin{subfigure}{0.3\textwidth}
		\includegraphics[width=\linewidth]{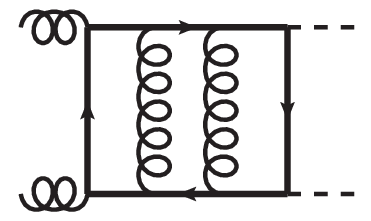}
		\captionsetup{justification=centering}
		\caption{\textit{One heavy quark loop:\\``$n_h$ diagrams''.}}
		\label{fig:dias:nh}
	\end{subfigure}
	\begin{subfigure}{0.3\textwidth}
		\includegraphics[width=\linewidth]{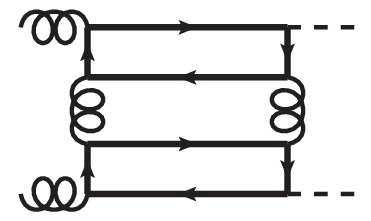}
		\captionsetup{justification=centering}
		\caption{\textit{Two heavy quark loops:\\``$n_h^2$ diagrams''.}}
		\label{fig:dias:nh2}
	\end{subfigure}
	\label{fig:dias}
	\caption{\textit{Example Feynman diagrams which contribute classes of diagrams with different fermion loops.}}
\end{figure}

In this section we define the notation and conventions used in these proceedings.
We study the process $g(q_1) g_(q_2) \to H(q_3) H(q_4)$, where all momenta are incoming; momentum conservation
implies that $q_4 = - q_1 - q_2 - q_3$, the Mandelstam variables are given by $s=(q_1+q_2)^2$, $t=(q_1+q_3)^2$
and $u=(q_1+q_4)^2$, and the transverse momentum of the Higgs bosons is given by
$p_T^2 = (u\,t - m_H^4)/s$.
The amplitude can be written as
\begin{equation}
	\mathcal{M}^{\mu\nu,ab} = \delta^{ab} X_0 s \left(F_1 A_1^{\mu\nu} + F_2 A_2^{\mu\nu} \right).
\end{equation}
Here, the external gluons' Lorentz indices are given by $\mu,\nu$ and adjoint colour indices by $a,b$.
$X_0 = \sqrt{2}\,G_F/2 \times T_F \alpha_s(\mu)/(2\pi)$ collects the overall factor, where $G_F$ is Fermi's constant,
$T_F=1/2$ and $\alpha_s(\mu)$ is the strong coupling constant at renormalization scale $\mu$. $A_1^{\mu\nu}$
and $A_2^{\mu\nu}$ are two linearly-indepenent Lorentz structures (which are not reproduced here),
whose coefficients are the so-called
``form factors'' $F_1$ and $F_2$. It is these form factors that we wish to compute as a perturbative expansion
in the strong coupling constant, with expansion coefficients defined as
$F = F^{(0)} + \left(\alpha_s(\mu)/\pi\right)F^{(1)} + \left(\alpha_s(\mu)/\pi\right)^2 F^{(2)} + \cdots$.
$F_1$ can be decomposed into a ``triangle'' part (which depends on $\lambda$) and a ``box'' part, where
both Higgs bosons couple separately to the top quark loop. Here we are only concerned with the ``box''
contributions. Furthermore we only consider one-particle irreducible diagrams; the reducible contribution
has been computed with different methods in Ref.~\cite{Davies:2024znp}. In these first steps
we concentrate on the classes of diagrams which
contain a light quark loop and a heavy quark loop (``$n_l$ diagrams'') (see Ref.~\cite{Davies:2023obx}) such as Fig.~\ref{fig:dias:nl},
or just one heavy quark loop (``$n_h$ diagrams'') as in Fig.~\ref{fig:dias:nh}.
Diagrams with two heavy quark loops (``$n_h^2$ diagrams'') such as in Fig.~\ref{fig:dias:nh2}, are left
for future work; the small-$t$ expansion of some of these diagrams is complicated by the presence of a massless $t$-channel cut.

\section{Small-$t$ Expansion}
\label{sec:smallt}
We begin by generating the diagrams which contribute to the three-loop amplitude with \texttt{qgraf}
\cite{Nogueira:1991ex}, which are mapped to integral topologies and written in a \texttt{FORM}
\cite{Ruijl:2017dtg} compatible notation with \texttt{tapir} \cite{Gerlach:2022qnc} and \texttt{exp}
\cite{Harlander:1998cmq,Seidensticker:1999bb}. The diagrams are then computed with the \texttt{FORM}-based
\texttt{calc} setup.

The loop integrals are first expanded around $m_H \to 0$ using \texttt{LiteRed} \cite{Lee:2013mka}.
The small-$t$ expansion is then realized in \texttt{FORM} by introducing $q_\delta = q_1+q_3$ (so $q_\delta^2 = t$)
in the loop integral propagators and expanding around $q_\delta \to 0$. After reducing the tensor
integrals which are produced by this expansion back to scalar integrals, a partial fraction decomposition
is performed with \texttt{tapir} or \texttt{LIMIT} \cite{Herren:2020ccq} to produce an expression in
terms of loop integrals without any linearly-dependent propagators. In this work at NNLO we compute only the
leading term in these expansions, i.e.~the final result contains only the $m_H^0 t^0$ term.

We now simplify the amplitude by mapping the loop integrals into a minimal set of topologies before
we consider the integration-by-parts (IBP) reduction. These maps are obtained by using \texttt{LiteRed}
or \texttt{Feynson} \cite{Maheria:2022dsq} to identify common sectors and sub-sectors between the topologies
of the amplitude. As a result the 60 topologies of the $n_l$ diagrams are written in terms of 28 topologies,
and the 522 topologies of the $n_h$ diagrams in terms of 203 topologies (for this set only \texttt{Feynson} is fast
enough to derive the maps).

\section{Integration-by-parts Reduction}
The next task is to express the loop integrals of the amplitudes in terms of a minimal set of ``master
integrals'', for which we use \texttt{Kira} \cite{Klappert:2020nbg}. For the $n_l$ diagrams the reduction
is relatively straightforward; the most complicated integral topology required around 1 week of runtime using
16 cores and consumed 500GB of memory. Symmetrizing the master integrals across all 28 topologies
with \texttt{Kira} results in a final reduction in terms of 177 master integrals.\footnote{We found one additional
relation between the master integrals, missed by \texttt{Kira}, by comparing amplitudes computed with two
different topology minimization mappings.}

For the $n_h$ diagrams the IBP reduction is much more complicated. The total running time for the 203
integral topologies was 330 days of 16 core jobs; the most difficult single topology required 41 days and
consumed more than 2TB of memory. Several optimizations were required for the reduction of this topology
to complete: a change of master integral basis was made in order to obtain $\epsilon$-factorizing
denominators in the reduction rules (using the \texttt{ImproveMasters.m} script which is included with
\texttt{FIRE} \cite{Smirnov:2023yhb}) and the routing of the loop momenta through the propagators was
adjusted compared to our initial definition such that each loop momentum takes the shortest path
through the graph.

The resulting reduction rules depend on 33K master integrals across all 203 topologies. Unlike in the
$n_l$ case we can not use a \texttt{Kira} reduction run across all topologies to reduce
these master integrals into a minimal set, as this would take much too long. We follow a more manual
set of steps:
\begin{enumerate}
	\item The \texttt{FindRules} function of \texttt{FIRE} maps the 33K master integrals to 4313.
	\item Apply \texttt{FindRules} to the full list of 2.6M integrals required by the
		amplitude.\footnote{For this purpose, we have improved \texttt{FindRules} such that it runs in parallel.}
		This reveals 1.3M pairs of equal integrals. Applying the IBP tables yields 820K non-trivial relations
		between 4029 of the master integrals. We solve these relations with the \texttt{user\_defined\_system}
		mode of \texttt{Kira} + \texttt{FireFly} \cite{Klappert:2019emp,Klappert:2020aqs}. The result is
		a reduction in terms of 1647 master integrals.
\end{enumerate}
The differential equation w.r.t.~$s$ for this basis of master integrals contains coupled blocks involving
integrals with different numbers of propagators, implying the basis is not minimal. We proceed to find
further relations as follows:
\begin{enumerate}
	\setcounter{enumi}{2}
	\item Perform a reduction of lists of ``trial integrals'' for each topology with \texttt{FIRE}. These
		reductions have a different basis of master integrals compared to that chosen by \texttt{Kira}.
		Repeating steps 1.~and 2.~above provides maps from 35K to 1817 and then to 1561 master integrals, which we map into a subset of the basis chosen by \texttt{Kira}.
\end{enumerate}
In this resulting basis the differential equation is simplified sufficiently that we can attempt to solve
it, though there is still no guarantee that this basis is fully minimal.

The differential equations for the master integrals are solved using the ``expand and match'' method
developed in Refs.~\cite{Fael:2021kyg,Fael:2021xdp,Fael:2022miw}. This provides ``semi-analytic'' results
by transporting analytic boundary conditions (from the point $s/m_t^2 = 0$, determined by large-mass
expansion and evaluation of tadpole integrals with \texttt{MATAD} \cite{Steinhauser:2000ry}) through a sequence of expansion points across the kinematic region by matching each expansion
numerically with its neighbour at a point at which they both converge. The $n_l$ master integrals have
all been determined in this way, and the $n_h$ master integrals are a work-in-progress, first for the
``large-$n_c$ subset'' of 783 integrals. We have verified that the $s/m_t^2 = 0$ limit of the $n_l$
and the large-$n_c$ $n_h$ amplitudes agree with the $t\to 0$ limit of
Ref.~\cite{Davies:2019djw}.

\section{Results}

\begin{figure}[t]
	\includegraphics[width=0.5\linewidth]{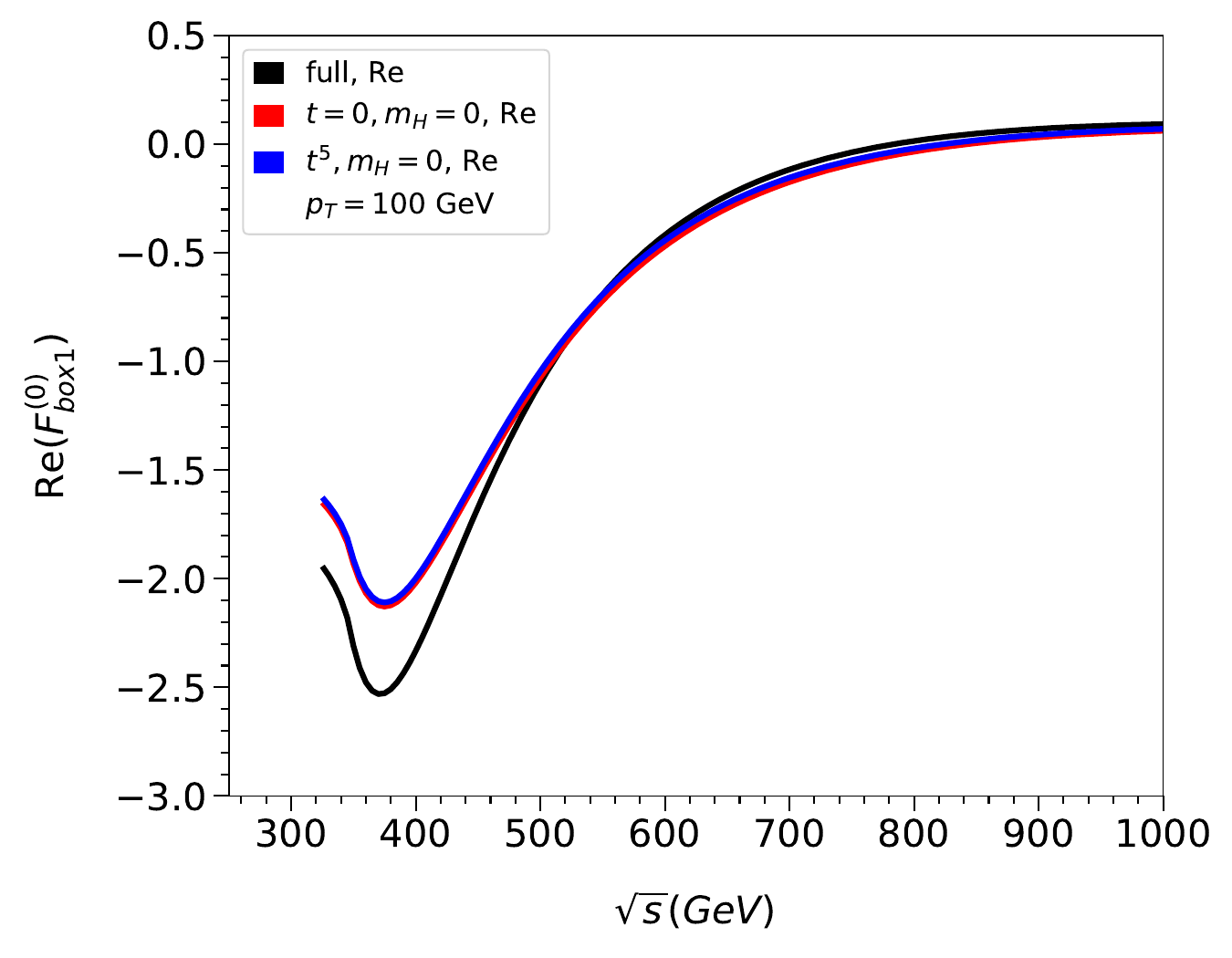}
	\includegraphics[width=0.5\linewidth]{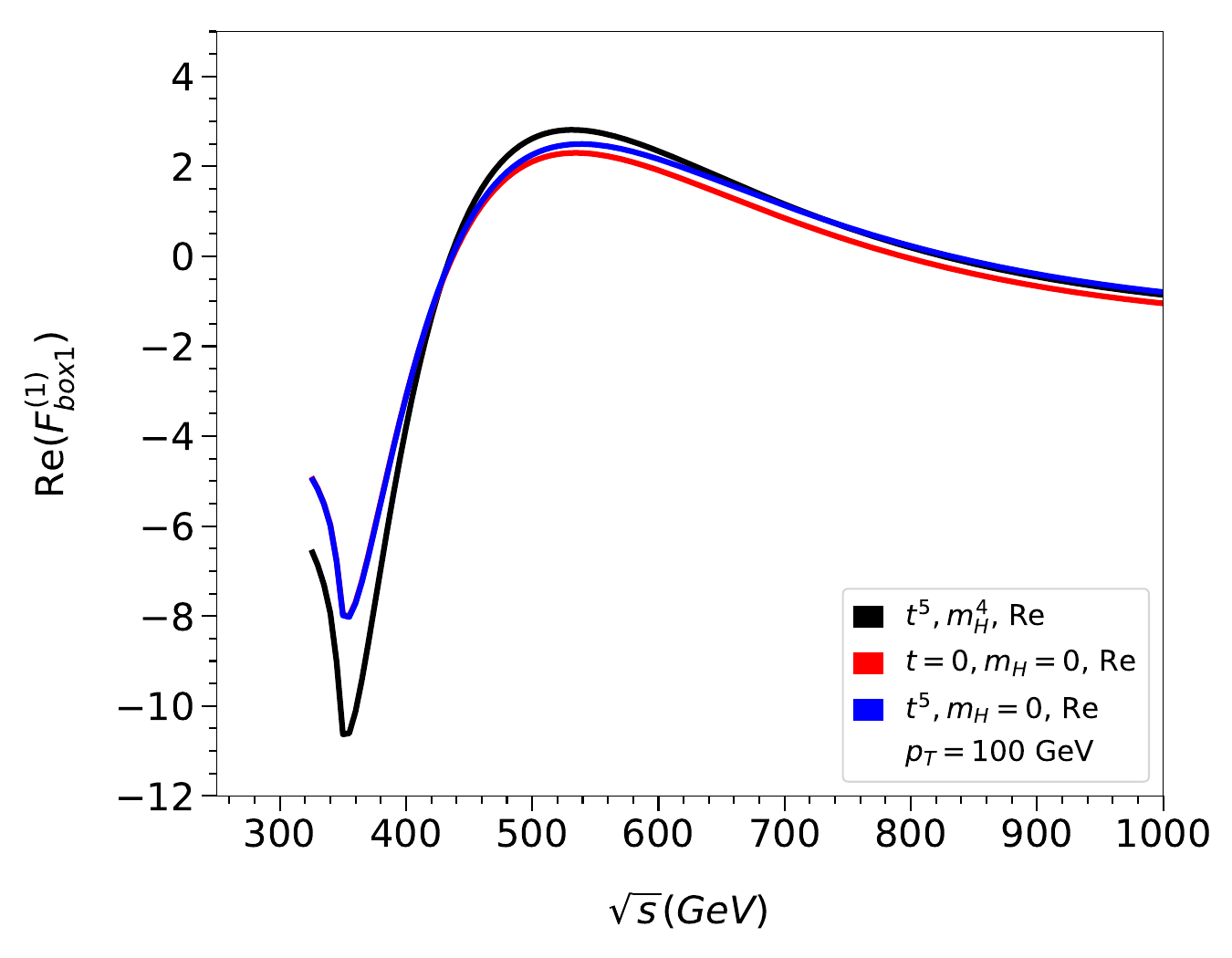}
	\caption{\textit{The ``box'' part of $F_1$ at one (left) and two (right) loops, as a function of $\sqrt{s}$
		with $p_T = 100$ GeV. The $m_H^0 t^0$ approximation is plotted in red and the $m_H^0 t^5$ approximation in
		blue. At one loop the exact curve with full dependence on $m_H, t$ is shown in black. At two loops we show
		the $m_H^4 t^5$ curve in black, since an exact result is not available.
	}}
	\label{fig:smallt-1l-2l}
\end{figure}

In Fig.~\ref{fig:smallt-1l-2l} we compare the results of the expansion described in Section \ref{sec:smallt}
with the exact result at one loop and with a deeper expansion at two loops, for which an exact result is
not available. 
We show curves for the form factor $F_1$ only, since $F_2$ vanishes at the leading expansion depth which
we study at NNLO.
We see that the $m_H^0 t^0$ approximation (in red) reproduces
the ``exact'' curves with around a 30\% accuracy for $\sqrt{s} < 500$ GeV. For higher energies this approximation
performs better. The blue curves show an expansion to $m_H^0 t^5$ and sit very close to the red curves, showing
that additional terms in the $m_H$ expansion are more important than additional $t$ terms for the plotted
value of $p_T=100$ GeV. For larger $p_T$ values the $t$ expansion terms become more important.

\begin{figure}[b]
	\centering
	\includegraphics[width=0.49\linewidth]{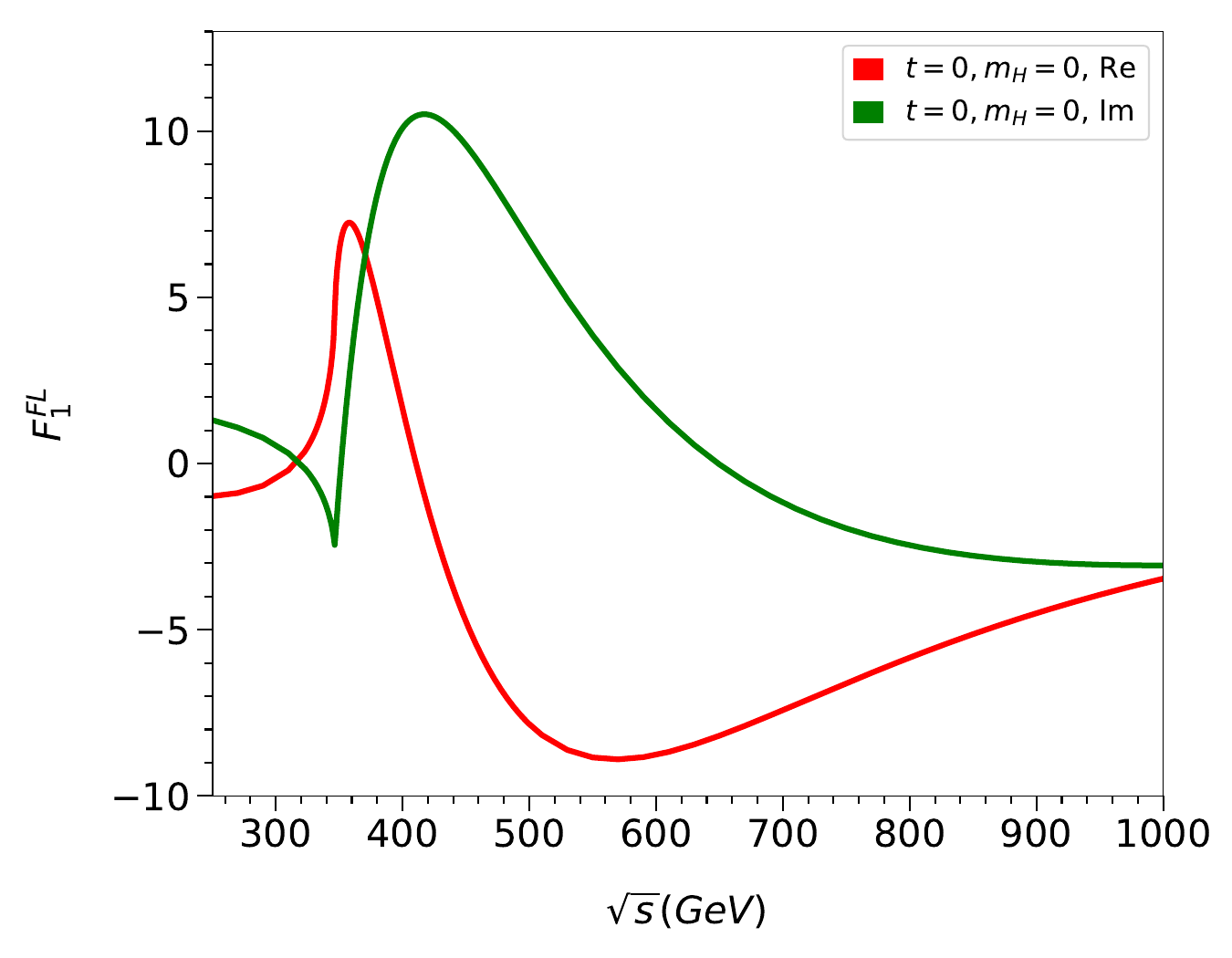}
	\includegraphics[width=0.49\linewidth]{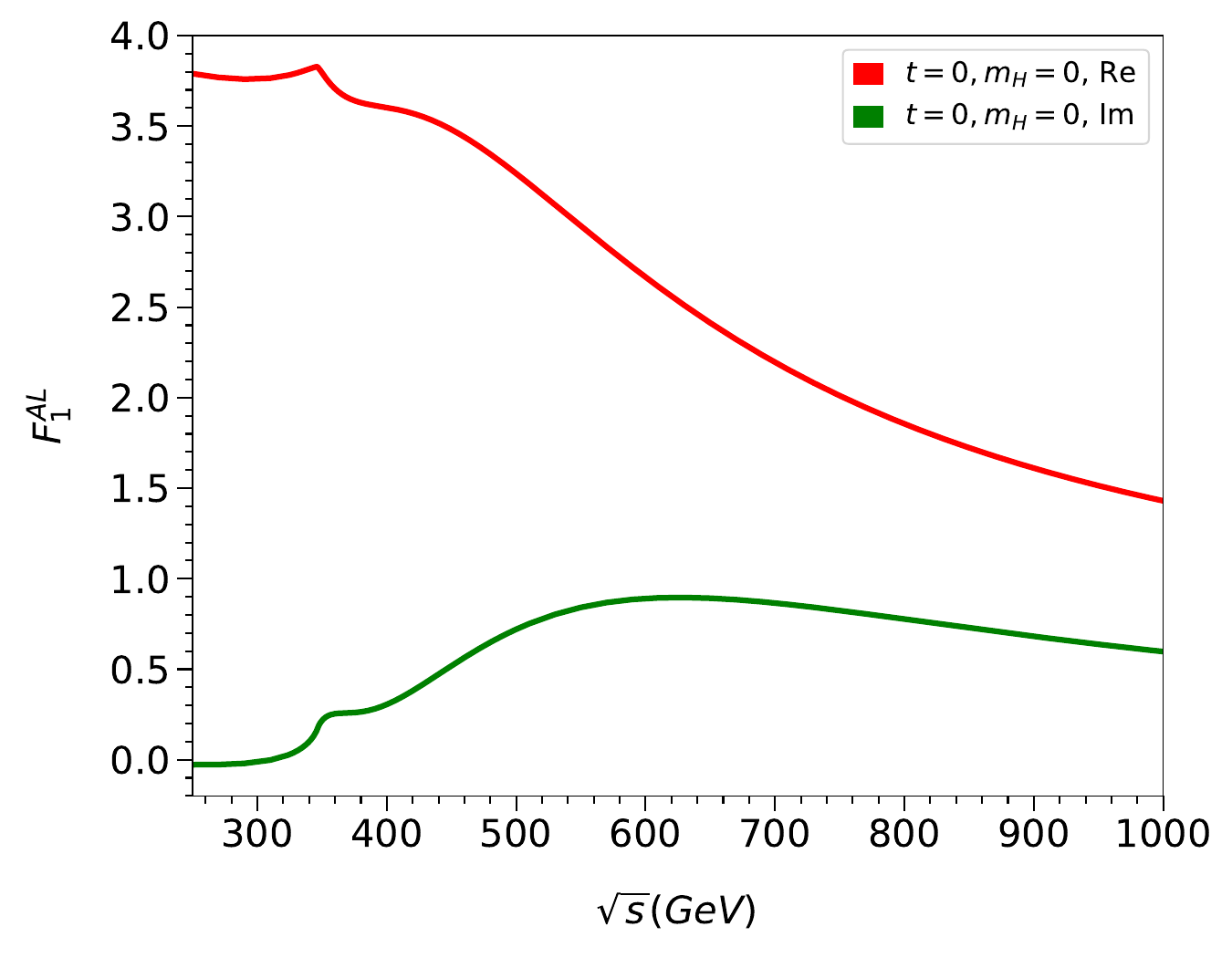}
	\includegraphics[width=0.49\linewidth]{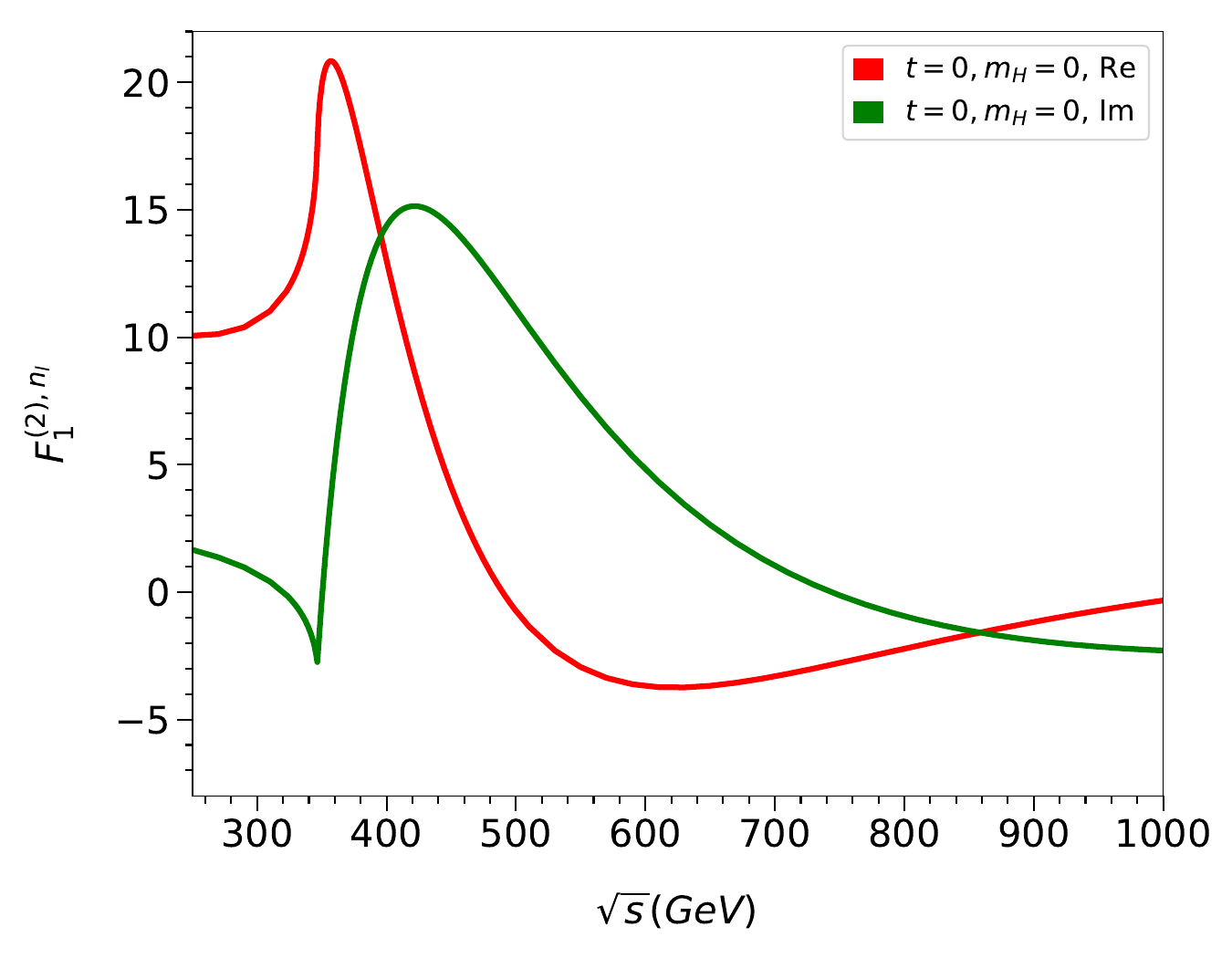}
	\caption{\textit{The $C_F$ (left) and $C_A$ (right) contributions to the $n_l$ diagram set, and their
		combination (lower). The real (red)
		and imaginary (green) parts are shown as a function of $\sqrt{s}$ for the leading term in the $m_H$ and
		small-$t$ expansion.
	}}
	\label{fig:smallt-3l-nl}
\end{figure}

Fig.~\ref{fig:smallt-3l-nl} shows the $m_H^0 t^0$ approximation for the $n_l$ diagrams, including both the
real (red) and imaginary (green) parts. The $C_F$ (left) and $C_A$ (right) contributions\footnote{$C_A$ and
$C_F$ are the adjoint and fundamental quadratic Casimir operators of $SU(3)$.} are plotted
separately, along with their combination (lower). Particularly for the $C_F$ part we observe a stronger variation
around the $2 m_t$ threshold
compared to the two-loop result; this is due to a leading logarithmic contribution of $v \log^2\!v$ at
three loops, compared to $v \log v$ at two loops, where $v = (1-4m_t^2/s)^{1/2}$.

\section{Outlook}
In these proceedings we have discussed progress towards a computation of the NNLO virtual amplitude
for Higgs boson pair production in gluon fusion. We have made use of an expansion in the forward
scattering limit in order to make the IBP reduction required by the amplitude feasible. The leading
term in the expansion approximates the full results with an error of about 30\% at LO and NLO.
Results are available \cite{Davies:2023obx} for the ``$n_l$ diagrams'' and the ``$n_h$ diagrams'' are
a work in progress. These diagram sets contain all necessary contributions to study the renormalization
scale and scheme dependence of the NNLO amplitude.
The $n_h^2$ diagrams will be computed in future work.

Additionally we aim to compute sub-leading terms in the expansion to improve quality of the approximation
and provide a non-zero value of the $F_2$ form factor. This will involve larger IBP reduction challenges
than we have considered at this stage, since the deeper expansion terms produce an amplitude which
depends on a larger number of Feynman integrals.

\section*{Acknowledgements}
The work of JD is supported by the STFC Consolidated Grant ST/X000699/1.

\newpage
\bibliographystyle{JHEP}
\bibliography{ll24-arxiv.bib}

\end{document}